\documentclass[conference]{IEEEtran}
\IEEEoverridecommandlockouts
\usepackage{cite}
\usepackage{amsmath,amssymb,amsfonts}
\usepackage{algorithmic}
\usepackage{graphicx}
\usepackage{textcomp}
\usepackage{xcolor}
\usepackage{subfigure}
\usepackage[ruled]{algorithm2e}
\usepackage{multirow}
\usepackage{cases}
\usepackage{booktabs}
\usepackage{url}
\def\BibTeX{{\rm B\kern-.05em{\sc i\kern-.025em b}\kern-.08em
    T\kern-.1667em\lower.7ex\hbox{E}\kern-.125emX}}
\begin{document}



\title{Demo: Paving the Way for Smart Manufacturing with 5G/TSN Convergence and Augmented Reality}

\author{
\IEEEauthorblockN{Sajida Gufran and Adnan Aijaz}
\IEEEauthorblockA{
\text{Bristol Research and Innovation Laboratory, Toshiba Europe Ltd., Bristol, United Kingdom}\\
\{firstname.lastname\}@toshiba-bril.com}
}    

\maketitle

\begin{abstract}
The fifth-generation (5G) mobile/cellular and time-sensitive networking (TSN)  \textcolor{black}{technologies are widely recognized as the key to shaping smart manufacturing for Industry 4.0 and beyond. Converged operation of the two offers end-to-end real-time and deterministic connectivity over hybrid wired and wireless segments. On the other hand, the augmented reality (AR) technology provides various benefits for the manufacturing sector. To this end, this demonstration showcases AR-aided remote assistance use-case over a hybrid TSN and 5G system. The demonstration setup comprises off-the-shelf 5G and TSN devices, a near product-grade 5G system, and an AR solution based on smart glasses. The demonstration shows the viability of over-the-air transmission of scheduled TSN traffic and real-time assistance for a local user from a remote environments. Performance results from the demonstration setup are also shown.} 

\end{abstract}


\section{Introduction}
\subsection{Background and Motivation}
\textcolor{black}{The global smart manufacturing market\footnote{https://www.marketsandmarkets.com/Market-Reports/smart-manufacturing-market-105448439.html} is expected to reach USD 240B by 2028. Smart manufacturing can be described as a collaborative and integrated production system, capable of responding in real time to changing demands. It resonates with the Industry 4.0 as well as the Industry 5.0 frameworks \cite{I4_I5}. The manufacturing industry is increasingly adopting the augmented reality (AR) technology as it improves operational efficiencies. AR-aided remote assistance can bridge the gap between field workers and experts. It improves workforce productivity, reduces production downtime, eases access to data for analytics, and simplifies training. }

Time-sensitive Networking (TSN)  provides deterministic, low-latency, and highly reliable communication over standard Ethernet networks~\cite{tsn}. With its precise timing and synchronization capabilities, TSN is poised to revolutionize various industries, including manufacturing, automotive, and energy. 
The IEEE 802.1Qbv TSN standard introduces time-aware scheduling, allowing for the reservation of time slots for critical traffic \cite{TSN_standards}. It also ensures that high-priority traffic is transmitted at specific times without interference from best-effort traffic. These features of IEEE 802.1Qbv are particularly promising for industrial automation.

The fifth generation (5G) mobile/cellular technology is designed to provide faster speeds, lower latency, and higher capacity. The ultra-reliable low-latency communication (uRLLC) capability of 5G provides millisecond-level latency which is crucial for real-time applications such as autonomous driving, and remote surgery, and smart manufacturing. With network slicing, it ensures dedicated resources and performance guarantees for different services \cite{demo_slicing_pvt_5g}. 

Integrating TSN and 5G leverages the strengths of both technologies, providing ultra-reliable, low-latency communication as well as timely and reliable data transmission over hybrid wired/wireless segments. It also supports applications requiring both high mobility and precise timing, typically found in smart manufacturing. Integration and converged operation of the two technologies is an important step in the evolution of industrial networks. From the perspective of Industry 4.0, such integration provides design simplifications by flattening the automation pyramid. From the perspective of Industry 5.0, it enables human-centric applications over converged wired and wireless segments, paving the way for  advanced manufacturing operations.


\subsection{Demonstration Overview and Distinguishing Aspects}
To this end, this demonstration aims at showcasing the benefits of integrating 5G and TSN technologies for smart manufacturing through AR-aided remote assistance use-case. The demonstration is based on our end-to-end hybrid 5G and TSN setup, which been used for evaluating IEEE 802.1Qbv scheduled traffic \cite{aijaz2024time}, and off-the-shelf TSN, 5G, and AR devices, along with a remote assistance solution. We demonstrate real-time operation of AR-aided remote assistance between a local user and a remote expert over a hybrid 5G and TSN system, along with end-to-end latency performance of the system.    

Prior demonstrations at similar venues have mainly shown the capabilities of private 5G, network slicing, or TSN features. To the best of our understanding, there has been no demonstration showing a real application over a hybrid wireless (5G) and wired (TSN) system. Demonstrating an AR-aided remote assistance use-case over 5G or TSN is a unique aspect as well.


\begin{figure*}
\centering
\includegraphics[scale=0.6]{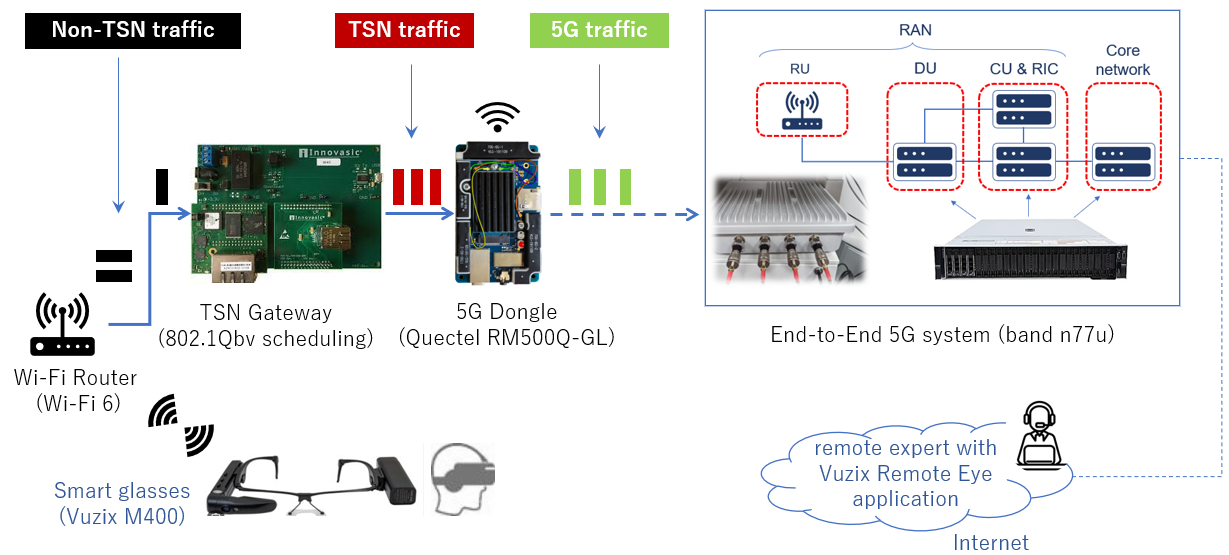}
\caption{Demonstration setup for AR-aided remote assistance use-case over a hybrid 5G and TSN system.}
\label{use_case_dia}
\end{figure*}




\section{Demonstration Setup}
\subsection{TSN Gateway}
We have used a TSN evaluation kit\footnote{RAPIDTSNEK-V0001} from Analog Devices in our setup. The TSN gateway allows any non-TSN device to participate in a TSN system without native implementation of TSN-specific features. The gateway functionality is provided between a
standard 100Base-TX Ethernet port and two 100Base-TX TSN ports. Further details of the TSN gateway and its configuration are available in ~\cite{aijaz2024time}.

\subsection{Smart Glasses}
We have used the M400 smart glasses equipment\footnote{https://www.vuzix.com/products/m400-smart-glasses} from Vuzix which provide lightweight, wearable, and ergonomically designed sensory and computing capabilities. 

\subsection{Remote Assistance Solution}
\begin{figure}
\centering
\includegraphics[scale=0.2]{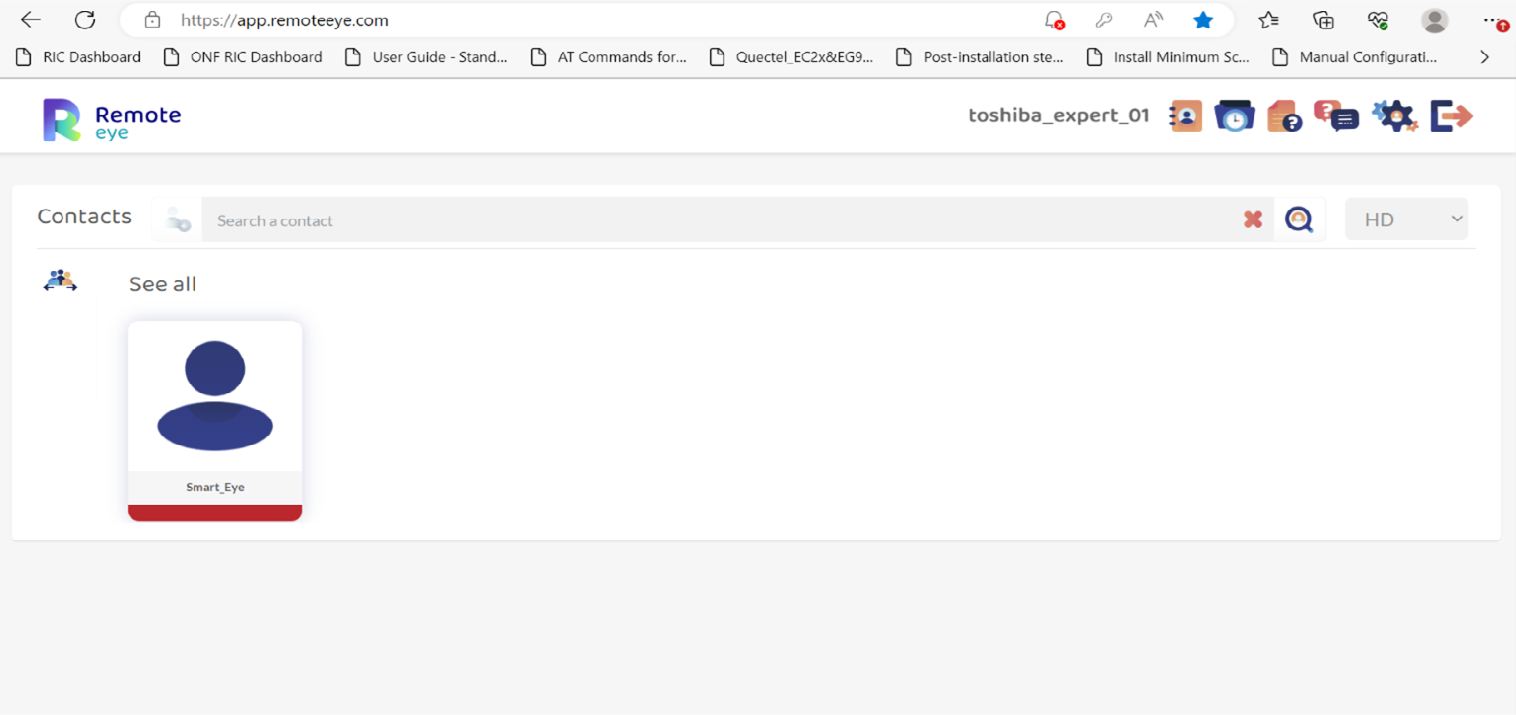}
\caption{Vuzix Remote Eye application.}
\label{rem_eye_dia}
\end{figure}

Our remote assistance solution is the  Vuzix Remote Eye, which is an enterprise-grade AR remote support and collaboration solution that leverages Vuzix's smart glasses and other wearable devices.  
This application is designed to facilitate real-time communication and assistance, enabling field workers and remote experts to collaborate effectively. A remote user can login and access the view on Vuzix glasses using remote eye application on any web server as shown in Fig.~\ref{rem_eye_dia}.


\subsection{5G System and Device}
The 5G system in our setup is an integrated standalone system based on O-RAN architecture and framework. It includes a radio unit (RU) and three general-purpose high-specification servers, running distributed unit (DU), centralized unit (CU), and 5G core network stacks. The suppliers of this system are listed as follows.

\begin{itemize}
    \item \textbf{RU} - Benetel (outdoor unit;
    RAN650\footnote{\url{https://benetel.com/ran650/}})
    \item \textbf{DU} - Effnet/Phluido 
    \item \textbf{CU} - Accelleran stack
    \item \textbf{Core network} - Attocore stack
\end{itemize}

We have used a Quectel RM500Q-GL module as the 5G device for this demo. Quectel RM500Q-GL is a 5G module specially optimized  for broadband applications. 

\subsection{Wi-Fi router}
We use a Wi-Fi router (with Wi-Fi 6 capabilities) for the smart glasses equipment to connect to the setup. The router is cabled to the non-TSN port of TSN gateway. 

\section{Demonstration Scenario and Key Results}
The demonstration use-case targets the smart manufacturing sector and aims to provide remote assistance/training to professionals for increased efficiency, productivity, and flexibility in manufacturing operations. We present a local user, wearing the smart glasses equipment which is connected to the wireless access point that is further connected to the non-TSN port of TSN gateway. Therefore, the traffic coming from the glasses is initially converted to the TSN traffic (based on IEEE 802.1Qbv). There is a 5G UE connected to the TSN port of the gateway and it relays the TSN traffic to the outside world through the 5G system. Once the 5G UE is connected to the Internet (through PDU session in the 5G system), the remote user can access to the glasses view using remote eye application and guides the local user.

The remote expert has access to the view of the smart glasses equipment which can be used to direct the local user. This scenario depicts human-aided real-time coordination in a manufacturing environment (e.g., production lines) improving production efficiency and quality. 

\begin{figure}
\centering
\includegraphics[scale=0.28]{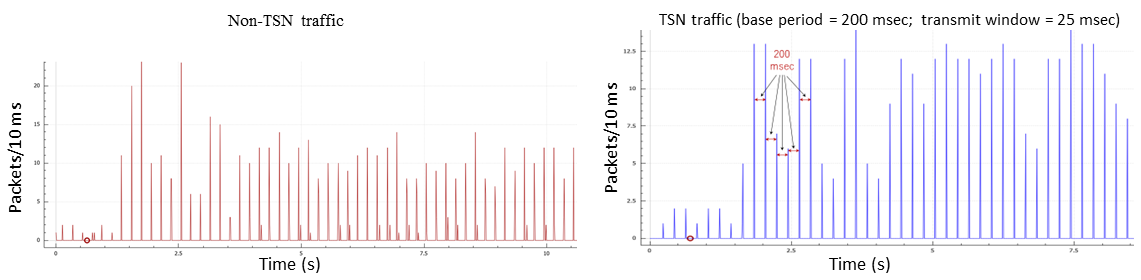}
\caption{Non-TSN and TSN traffic.}
\label{non_tsn_and_tsn_traffic}
\end{figure}

The conversion of non-TSN to TSN traffic is highlighted in Fig.~\ref{non_tsn_and_tsn_traffic}. 
Further details about the behavior of TSN traffic under different parametric settings are covered in our recent work ~\cite{aijaz2024time}

\begin{figure}
\centering
\includegraphics[scale=0.35]{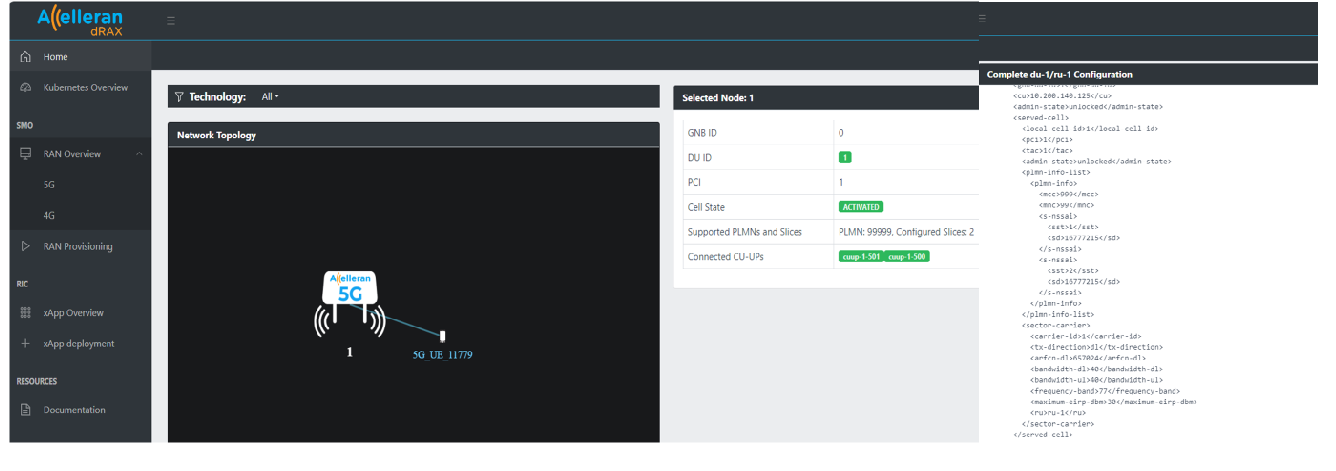}
\caption{5G RAN dashboard (showing successful UE connectivity).}
\label{ric_dash}
\end{figure}

\begin{figure}
\centering
\includegraphics[scale=0.4]{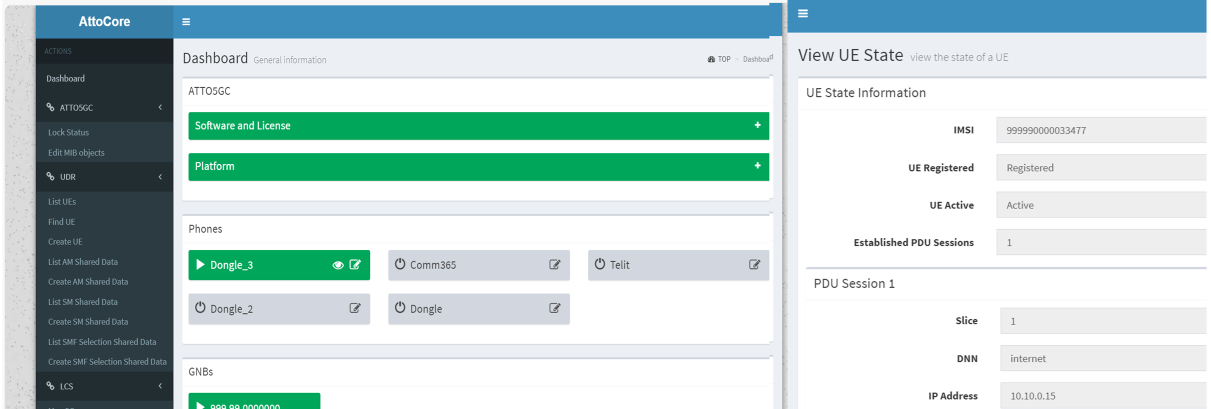}
\caption{5G core network dashboard (showing successful data connectivity).}
\label{5g_core_dash}
\end{figure}

From the 5G system perspective, the UE status is shown in the RAN and core dashboards, in Fig. \ref{ric_dash} and Fig. \ref{5g_core_dash}, respectively. From 5G core dashboard, we can also see the PDU session details including the allocated network slice information. The slice information in RAN can also be viewed from RAN dashboard Fig.~\ref{ric_dash}.


\begin{figure}
\centering
\includegraphics[scale=0.45]{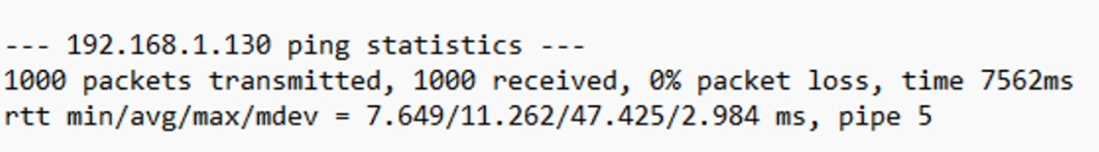}
\caption{End-to-End latency between the TSN gateway (non-TSN port) and the 5G system (core network).}
\label{delay_dia}
\end{figure}

Next, we evaluate the latency performance. 
The measured round-trip latency between the non-TSN port of the TSN gateway and the 5G core network is 11 msec on average, which is shown in Fig.~\ref{delay_dia}.

\begin{figure}
\centering
\includegraphics[width=\columnwidth]{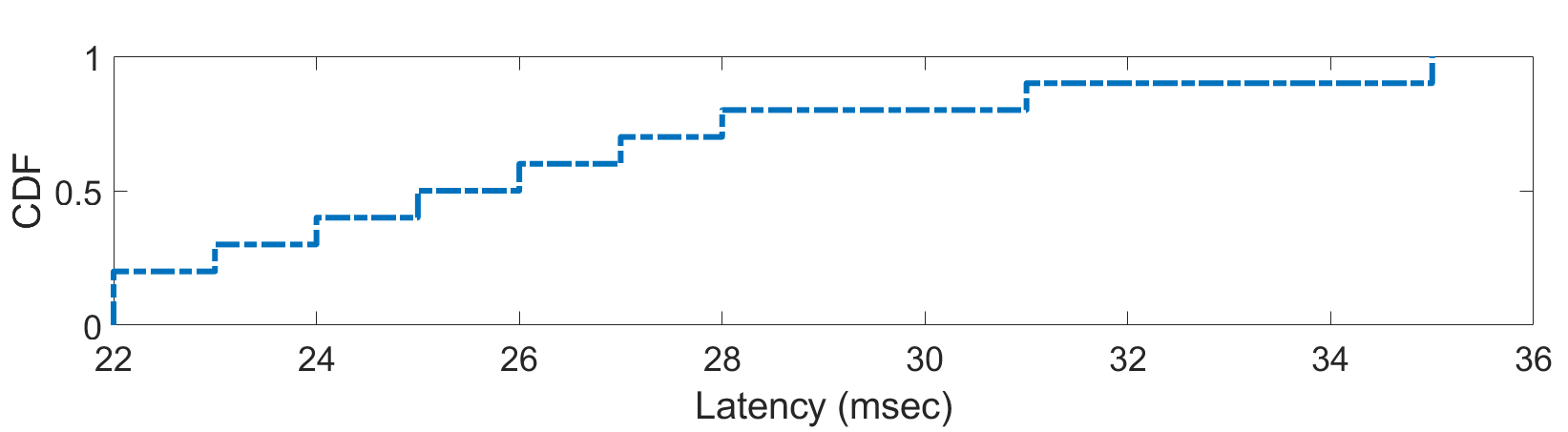}
\caption{CDF plot of the latency between the local user and the remote expert.}
\label{user_delay_dia}
\end{figure}

With the help of clock zone\footnote{\url{https://clock.zone/}}, we have calculated the latency between local user and remote user. We took multiple readings, revealing latency of 22 to 35 msec as per the cumulative distribution function (CDF) plot in Fig.~\ref{user_delay_dia}.  This latency takes into account latency in the Wi-Fi connection (5 to 6 ms), 10 to 12 ms in between non-TSN port and 5G core, as well as additional latency in the cable and in the Internet.

The Remote Eye application has been developed to enhance collaboration, productivity, and safety across various industries by enabling real-time, hands-free communication and support. Using this application with 5G and TSN technologies, makes it highly reliable and responsive. Experts can guide technicians through complex repairs in real-time, as if they were on-site. The 5G broadband capabilities support advanced AR functionalities, enhancing the visual guidance provided through the application, while TSN enables guaranteed data delivery. In our demonstration, the end users communicate seamlessly with each other. Multiple users can connect simultaneously and get real-time fault correction. They can upload and share the error logs with remote user; the high bandwidth and low latency capabilities make this information exchange seamless to the end users.

\section{Remarks}\label{sect_cr}
This demonstration highlighted the value of 5G and TSN convergence and AR-aided remote assistance for smart manufacturing operations. The real-time, broadband, and deterministic connectivity is important for manufacturing processes. It also enables advanced automation, integration of artificial intelligence, and human-in-the-loop, paving the way for agile, responsive, and intelligent manufacturing environments. Even though the demonstration focused on smart manufacturing, the results and insights are applicable to other industry verticals, e.g., in healthcare. 
A video of the demonstration is available at \url{https://tinyurl.com/ywtdswec}.

\section*{Acknowledgements}
The 5G Open RAN setup used in this demonstration was partly supported by the DSIT (formerly DCMS) of UK government as part of the FRANC programme.

\bibliographystyle{IEEEtran}
\bibliography{bibliography.bib}
\end{document}